\documentclass[a4paper]{article}
\usepackage{geometry}
\usepackage{setspace}
\usepackage{float}
\usepackage{epstopdf}
\usepackage{standalone}
\usepackage{amsmath}          
\newtheorem{assumption}{Assumption}     
\newtheorem{theorem}{Theorem}    
\newtheorem{corollary}{Corollary} 
\usepackage{bm}
\usepackage{pdflscape}
\usepackage{authblk}
\usepackage{graphicx}
\usepackage[flushleft]{threeparttable}
\usepackage[comma,authoryear]{natbib}
\usepackage[citecolor=blue]{hyperref}

\usepackage{mathtools}
\DeclarePairedDelimiter\ceil{\lceil}{\rceil}

\newcommand{\interior}[1]{%
 {\kern0pt#1}^{\mathrm{o}}%
}
\usepackage [english]{babel}
\usepackage [autostyle, english = american]{csquotes}
\MakeOuterQuote{"}

\usepackage{algorithm,algpseudocode}
\usepackage{caption}
\usepackage[makeroom]{cancel}

\title{\textbf{A Function Emulation Approach for Doubly Intractable Distributions}}
\author[1]{Jaewoo Park}
\author[1]{Murali Haran}
\affil[1]{Department of Statistics, The Pennsylvania State University}

\begin{document}

\maketitle

\begin{abstract}
Doubly intractable distributions arise in many settings, for example in Markov models for point processes and exponential random graph models for networks. Bayesian inference for these models is challenging because they involve intractable normalising "constants" that are actually functions of the parameters of interest.  Although several computational methods have been developed for these models, each can be computationally burdensome or even infeasible for many problems. We propose a novel algorithm that provides computational gains over existing methods by replacing Monte Carlo approximations to the normalising function with a Gaussian process-based approximation. We provide theoretical justification for this method. We also develop a closely related algorithm that is applicable more broadly to any likelihood function that is expensive to evaluate. We illustrate the application of our methods to challenging simulated and real data examples, including an exponential random graph model, a Markov point process, and a model for infectious disease dynamics. The algorithm shows significant gains in computational efficiency over existing methods, and has the potential for greater gains for more challenging problems. For a random graph model example, we show how this gain in efficiency allows us to carry out accurate Bayesian inference when other algorithms are computationally impractical.
\end{abstract}

\noindent%

{\it Keywords: Markov chain Monte Carlo; doubly intractable distributions; exponential random graph models; Markov point processes; importance sampling; Gaussian processes} 
\vfill

\section{Introduction}

~~~~~Models with intractable normalising functions arise frequently, for example in exponential random graph models \citep[cf.][]{robins2007introduction, hunter2012inference} for social networks, autologistic models \citep[cf.][for a review]{besag1974spatial, hughes2011autologistic} for lattice data, and interaction spatial point process models \citep[cf.][]{strauss1975model,goldstein2014attraction}. Consider $h(\mathbf{x}|\bm{\theta})$, an unnormalized probability model for a data set $\mathbf{x} \in \mathcal{X}$ given a parameter vector $\bm{\theta} \in \bm{\Theta}$. Suppose it has a normalising function $Z(\bm{\theta})=\int_{\mathcal{X}} h(\mathbf{x}|\bm{\theta})d\mathbf{x}$. Let $p(\bm{\theta})$ be the prior density for $\bm{\theta}$. The likelihood function, $L(\bm{\theta}|\mathbf{x})$ is $h(\mathbf{x}|\bm{\theta})/Z(\bm{\theta})$ and the posterior density of $\bm{\theta}$ is

\begin{equation}
\pi(\bm{\theta}|\mathbf{x}) \propto p(\bm{\theta})\frac{h(\mathbf{x}|\bm{\theta})}{Z(\bm{\theta})}.
\label{e1}
\end{equation}

\noindent In Bayesian analysis this results in so-called doubly intractable posterior distributions. The major computational issue for these models is that $Z(\bm{\theta})$ cannot be easily evaluated. Several algorithms substitute $Z(\bm{\theta})$ with a Monte Carlo approximation. However, such approximations are often computationally expensive, making the resulting Markov chain Monte Carlo (MCMC) algorithm impractical. In this manuscript we provide an approach for replacing Monte Carlo approximations with fast Gaussian process approximations. We demonstrate how this algorithm is fast while producing accurate posterior approximations. Later we  discuss the case where $h(\mathbf{x}|\bm{\theta})$ is very expensive to compute and propose a method to solve the general problem where the likelihood function is expensive to evaluate.

There is a large literature on computational methods for doubly intractable distributions. \cite{besag1974spatial} proposed the pseudolikelihood approximation, a simple approximation to $h(\mathbf{x}|\bm{\theta})$ that does not contain $Z(\bm{\theta})$. However in the presence of strong dependence among data points, the maximum pseudo-likelihood estimator (MPLE) can be a poor approximation to the MLE. \cite{geyer1992constrained} proposes MCMC-MLE which is based on maximizing a Monte Carlo approximation to the likelihood. This approach is elegant and practical, but the algorithm requires analytical gradients for the unnormalized likelihood, which is not available in many cases \citep[cf.][]{goldstein2014attraction}. Bayesian alternatives may be useful in such situations, and also in cases where we want a convenient approach for carrying out inference for hierarchical models involving normalising function models, and for  incorporating prior information. 
There is a growing literature on computational methods for Bayesian inference for such models \cite[see][for a review]{park2018bayesian}. Asymptotically exact algorithms are those where the Markov chain's stationary distribution is equal to the desired posterior distribution. Examples of elegant asymptotically exact methods for this challenging problem include \cite{moller2006efficient,murray2006,atchade2008bayesian,lyne2015russian,liang2015adaptive}. While some require the ability to draw independent samples from the probability model, others are complicated to construct and require users to tune the algorithm carefully. All of them are computationally infeasible for many interesting models \citep{park2018bayesian}. Asymptotically inexact approaches may be much faster   \citep{liang2010double,alquier2014noisy}, but they can  still be prohibitively expensive, for instance in the case of an exponential random graph model example we describe later in this manuscript. This motivates the development of computationally efficient algorithms that allow scientists to fit models to more complex models and larger data sets than previously possible. 

In this manuscript we describe an algorithm that uses a  fast two-stage approximation to construct an efficient MCMC algorithm for doubly intractable distributions. The two steps are as follows: (1) approximate the normalising function at several parameter values using importance sampling, and (2) interpolate the normalising function at other parameter values using a Gaussian process fit to the importance sampling approximations. These two steps allow the normalising function to be approximated when evaluating Metropolis-Hastings acceptance probabilities in a Markov chain Monte Carlo algorithm.  

Gaussian processes have been widely used for interpolation in spatial statistics  \citep{kbiob1951statistical,cressie2015statistics}, as well as in "computer model emulation", to approximate the relationship between input parameters  and the output of a complex computer model \cite[cf.][]{sacks1989design,kennedy2001bayesian}. We show how Gaussian processes are very effective in our two-stage approximation, and how our method may be useful in   addressing inferential challenges for doubly intractable distributions. We also describe a second algorithm that is applicable in principle to a much wider class of problems -- Bayesian inference when the likelihood function (not just its normalising function) is difficult to evaluate.

The outline of the remainder of this paper is as follows. In Section 2 we describe existing Bayesian algorithms for intractable normalising functions and discuss their computational challenges. We also introduce several function emulation approaches used in a number of works in the areas of computational statistics. In Section 3 we propose our fast Gaussian process-based function emulation approach to inference, and provide implementation details. In addition, we provide  theoretical justification for our algorithm. In Section 4 we describe the application of our approach in the context of three different case studies, including a more general problem where the entire likelihood function is assumed to be intractable. We study the computational and statistical efficiency of our algorithm, showing how our algorithm is able to perform inference for a problem where others are computationally impractical. We conclude with a summary and discussion in Section 5.

\section{Computational Methods}

\subsection{Bayesian Methods for Doubly Intractable Distributions}

~~~~~Several MCMC algorithms have been developed for Bayesian inference for doubly intractable distributions. \cite{park2018bayesian} classifies these algorithms into two broad if somewhat overlapping categories: (1) {\it{ likelihood approximation approaches }} which directly approximate  the normalising functions via importance sampling, and substitute the approximations into the Metropolis-Hastings acceptance probability \citep{atchade2008bayesian,lyne2015russian,alquier2014noisy}, and (2) {\it{auxiliary variable approaches}} which introduce an auxiliary variable that cancels out the normalising functions in the Metropolis-Hasings acceptance probability. 


For instance, \cite{lyne2015russian} constructs an unbiased likelihood estimate for doubly-intractable distributions. This method has the advantage that it can be shown to be a pseudo-marginal algorithm which is known to be asymptotically exact \citep{beaumont2003estimation,andrieu2009pseudo}. However, for the problems we consider in this paper, specifically doubly intractable distributions, obtaining a single likelihood estimate requires multiple Monte Carlo approximations of $Z(\bm{\theta})$. This is computationally infeasible even for small data problems involving the interaction point process and exponential random graph model we consider in this paper \citep{park2018bayesian}. \cite{moller2006efficient,murray2006} relies on perfect sampling \citep{propp1996exact}, an algorithm that uses bounding Markov chains to generate an auxiliary variable that is exactly from the target distribution. These algorithms are asympotically exact. However, perfect samplers are available only for a small set of probability models, and even for these cases, they tend to be very computationally expensive; this  greatly limits the applicability of algorithms that require perfect sampling. To address this, \cite{liang2010double} proposes a double Metropolis-Hastings (DMH) algorithm by replacing perfect sampling with a standard Metropolis-Hastings algorithm. Although DMH is asymptotically inexact, among current approaches it is the most practical method for computationally expensive problems (see \cite{park2018bayesian} for details). But even the DMH algorithm is  computationally infeasible in some situations, as we will show via examples in Section 4.

Approximate Bayesian computation (ABC) methods \citep{beaumont2002approximate,marin2012approximate} are popular likelihood-free algorithms. In their simplest form, for a given parameter drawn from the prior distribution, ABC methods simulate a representative summary statistic from the model. If this simulated summary statistic is close to the observed summary statistic, the parameter is accepted. Accepted samples are approximately distributed according to the posterior distribution. ABC and its MCMC variant \citep{marjoram2003markov} are broadly applicable. However, for models without representative summary statistics, ABC can be very inefficient. 
 
Current auxiliary variable and likelihood approximation algorithms are computationally expensive when the data are high-dimensional. The main expense is due to the high-dimensional auxiliary data simulations.  All the algorithms require sampling a data set ($\mathbf{x}$) from the probability model ($h(\mathbf{x}|\bm{\theta})$) at each iteration of the algorithm. Multiple samples are generated to construct an importance sampling estimate (likelihood approximation approach) or a single sample is simulated to cancel out $Z(\bm{\theta})$ (auxiliary variable approach). The sampling becomes more demanding as the dimension of the data ($\mathbf{x}$) increases. Furthermore,  adaptive algorithms \citep{atchade2008bayesian,liang2015adaptive} are computationally infeasible for high-dimensional data sets in some cases because in order to guarantee asymptotically exact inference, the adaptive algorithms require storing simulated auxiliary data with each iteration. For models without low-dimensional summary statistics, the memory costs can become prohibitively expensive.                                                        

\subsection{Function Emulation}

~~~~~In many disciplines including climate science, mechanical engineering, computer models are used to simulate complex processes. Because these numerical simulations are expensive, it becomes difficult to study how the processes vary as functions of parameters and it is challenging to perform statistical inference. Several global approximations for such models have been developed using polynomial functions \citep{marzouk2007stochastic,marzouk2009stochastic}, radial basis functions \citep{joseph2012bayesian,bliznyuk2012local}, Gaussian processes \citep{sacks1989design, kennedy2001bayesian, rasmussen2004gaussian, wang2017adaptive} and many others. However, analyzing convergence and the error of global approximations is often challenging. To overcome such difficulties of uniform modeling, local and nonstationary approximations are also studied in the Gaussian process context \citep{gramacy2008bayesian, gramacy2015local} and polynomial functions \citep{conrad2016accelerating}. For instance, \cite{conrad2016accelerating} replace likelihood functions with local polynomial approximations in the Metropolis-Hastings kernel. With increasing iterations, these local approximations are refined via sequential experimental design procedure, so that the approximations are asymptotically exact. This is an elegant approach but does not apply to our problem. They assume that it is possible to evaluate the likelihood exactly, even if each evaluation is expensive. In the doubly intractable distributions context, exact likelihood evaluations are not possible.  Our problem requires another layer of approximation to the intractable normalizing functions; this can lead to further computational difficulties. We note, however, that for our disease dynamics example in the supplementary material, a version of the algorithm in \cite{conrad2016accelerating} may be a possible alternative to our algorithm. 

Several methods based on Gaussian process approximation have been proposed to accelerate inference when the likelihood function is intractable or expensive. Gaussian process approximations have also been used in the ABC context \citep{wilkinson2014accelerating,meeds2014gps,gutmann2016bayesian,jarvenpaa2016gaussian}. For instance, \cite{wilkinson2014accelerating} uses a Gaussian process to reduce the number of simulations in ABC. This method can iteratively rule out implausible regions of the parameter space. We note that there are delayed-acceptance approaches to speed up MCMC algorithms using cheap surrogates \citep{christen2005markov,golightly2015delayed,sherlock2017adaptive}. By using cheap approximations, poor proposals are rejected quickly, and expensive likelihood functions are evaluated at the next stage only for promising proposals. Although delayed-acceptance MCMC approaches are asymptotically exact, the efficiency gains are limited because the methods require evaluating expensive likelihood functions at the second stage for promising proposals. 
 
 \cite{drovandi2018accelerating} proposes an approach to speed up pseudo-marginal methods by replacing the log of an unbiased likelihood estimate with a Gaussian process approximation. Our approach has similarities to \cite{drovandi2018accelerating} in that we replace the log of the function estimate with a Gaussian process approximation, and also use a short run of an MCMC algorithm to obtain good design points for constructing the Gaussian process approximation. Function emulation approaches such as \cite{drovandi2018accelerating} and the method we describe in this manuscript, are useful for problems where it is expensive to evaluate a function (or evaluate a function approximation) many times. Both our algorithm and the algorithms in \cite{drovandi2018accelerating} require a pre-computation step which itself involves repeated approximations of the likelihood function at a set of parameter values. \cite{drovandi2018accelerating} requires unbiased likelihood estimates in the pre-computation step, which can be prohibitively expensive for the problems we consider -- interaction point processes and exponential random graph models with large data sets. In contrast, our approach, as we demonstrate later in the paper, is computationally efficient even for such problems. To our knowledge, no existing approach provides general computer model emulation framework for doubly intractable distributions. We note that \cite{reich2014spatial} develops an approximate MCMC method for the Strauss process \citep{strauss1975model} using polynomial interpolation. However our approach applies more broadly due to the flexibility and nonparametric nature of the approach (the covariance mimics the role of a non-linear relationship), and we have  studied its application to several challenging examples. Furthermore, we are able to provide a theoretical justification for our methodology.

\section{Markov chain Monte Carlo Using Gaussian process-based Function Emulation}

~~~~~Here we describe two algorithms: NormEm  constructs a Markov chain Monte Carlo algorithm for  approximating a doubly intractable distribution, and  LikEm applies more broadly to posterior distributions where the entire likelihood function is hard to evaluate.

\subsection{Outline}

~~~~~Gaussian processes are commonly used for nonparametric regression \citep[cf.][]{rasmussen2004gaussian}, in spatial interpolation \citep[cf.][]{kbiob1951statistical,cressie2015statistics}, and in approximating computationally expensive computer models \citep{sacks1989design}. 
 The main idea of our approach is to replace expensive importance sampling estimates with fast Gaussian process approximations. We can approximate either $Z(\bm{\theta})$ (normalising function emulation) or $L(\bm{\theta}|\mathbf{x})$ (full likelihood function emulation). We begin with an  outline of the normalising function emulation algorithm. \\  

\noindent {\it{Step 1.}} Log of importance sampling estimates for $Z(\bm{\theta})$ are computed at a set of $\bm{\theta}$ values.\\
{\it{Step 2.}} A Gaussian process model is fit to the above estimates, which allows for approximation of $\log Z(\bm{\theta})$ at other $\bm{\theta}$ values.\\
{\it{Step 3.}} A Markov chain Monte Carlo algorithm is constructed for sampling from the posterior distribution of $\bm{\theta}$ where for each Metropolis-Hasting accept-reject ratio, the approximation from Step 2 is used.\\

Our second algorithm is similar but it directly approximates $\log L(\bm{\theta}|\mathbf{x})$ instead of approximating just  $\log Z(\bm{\theta})$. We provide details in the following section.

\subsection{Function Emulation Algorithms}

~~~~~For a $p$-dimensional parameter vector $\bm{\theta} \in \bm{\Theta}$, consider $d$ particles in $\bm{\Theta}$, $\bm{\psi} =( \bm{\theta}^{(1)},\dots,\bm{\theta}^{(d)} )'$. Let  $\bm{\widetilde{\theta}}$ be an approximation to the maximum likelihood estimate, for example, the maximum pseudolikelihood estimate \citep{besag1974spatial} or sample mean of $\bm{\psi}$. For $i=1,\dots,d$ we can construct unbiased Monte Carlo estimates for $( Z(\bm{\theta}^{(1)})/Z(\bm{\widetilde{\theta}}),\dots,Z(\bm{\theta}^{(d)})/Z(\bm{\widetilde{\theta}}) )$ via importance sampling. For each $i$, log of the importance sampling estimate is 
 
\begin{equation}
\log \widehat{Z}_{IS}(\bm{\theta}^{(i)})=\log \Big( \frac{1}{N}\sum_{l=1}^{N}\frac{h(\mathbf{x}_{l}|\bm{\theta}^{(i)})}{h(\mathbf{x}_{l}|\bm{\widetilde{\theta}})} \Big),
\label{IS}
\end{equation}

\noindent where each $\mathbf{x}_{l}$ is the last draw of the $l$th Markov chain with stationary distribution $h(\cdot|\bm{\widetilde{\theta}})/Z(\bm{\widetilde{\theta}})$. For a robust approximation, an importance sampling estimate can also be extended through umbrella sampling \citep{torrie1977nonphysical,atchade2008bayesian, geye:2011}. The log importance sampling approximations $\log \mathbf{\widehat{Z}_{IS}} = ( \log \widehat{Z}_{IS}(\bm{\theta}^{(1)}),\dots,\log \widehat{Z}_{IS}(\bm{\theta}^{(d)}) )' \in R^{d}$ are obtained, respectively, at the particles $\bm{\psi} =( \bm{\theta}^{(1)},\dots,\bm{\theta}^{(d)} )' \in R^{d \times p}$. We can construct a Gaussian process model relating the importance approximation to the particle, 

\begin{equation}
\log \mathbf{\widehat{Z}_{IS}}= \bm{\mu} + \mathbf{u}, 
\label{GP}
\end{equation}

\noindent where $\bm{\mu}$ is the mean and $\mathbf{u}$ is a second order stationary Gaussian process. For $i,j =1,\dots,d$, a symmetric and positive definite covariance function can be defined as

\begin{equation}
\mathbf{K}(\bm{\theta}^{(i)},\bm{\theta}^{(j)};\sigma^2,\phi,\tau^2) = \sigma^2\Big(  1 + \frac{\sqrt{3}\|\bm{\theta}^{(i)}-\bm{\theta}^{(j)}\|}{\phi} \Big) \exp{\Big( -\frac{\sqrt{3}\|\bm{\theta}^{(i)}-\bm{\theta}^{(j)}\|}{\phi}   \Big)} + \tau^2 1_{\lbrace i = j \rbrace},
\label{matern}
\end{equation}	
	
\noindent with partial sill $\sigma^2$, range $\phi$, and nugget $\tau^2$. Since we assume that the log-normalizing function surface is smooth, we take a Mat\'{e}rn class covariance function, where the smoothness parameter is set to $3/2$. We assume a simple linear mean trend $\bm{\mu}=\bm{\psi}\bm{\beta}$, where $\bm{\beta} \in R^{p}$ is the regression parameter. The flexible covariance structure allows, indirectly, for a "nonparametric" non-linear mean function; this is the basis for kriging and computer model emulation. 

To obtain $\log \widehat{Z}_{IS}(\bm{\theta}^{\ast})$ at some new $\bm{\theta}^{\ast}\in \Theta$, we use basic definitions of the Gaussian process to obtain

\begin{equation}
\begin{bmatrix}
\log \mathbf{\widehat{Z}_{IS}} \\
\log \widehat{Z}_{IS}(\bm{\theta}^{\ast})
\end{bmatrix}
=MVN\left( \begin{bmatrix}
     \bm{\psi}\bm{\beta} \\
     \bm{\theta}^{\ast}\bm{\beta} 
     \end{bmatrix}
   ,
  \begin{bmatrix}
    \mathbf{C} & \mathbf{c} \\
    \mathbf{c}' & \sigma^2+\tau^2
  \end{bmatrix}
  \right),
\label{GPmat}
\end{equation}

\noindent where $\mathbf{C}=\mathbf{K}(\bm{\psi},\bm{\psi};\sigma^2,\phi) \in R^{d \times d}$ and $\mathbf{c}=\mathbf{K}(\bm{\psi},\bm{\theta}^{\ast};\sigma^2,\phi) \in R^{d \times 1}$. The conditional distribution of $\log \widehat{Z}_{IS}(\bm{\theta}^{\ast})$ given  observed $\log \mathbf{\widehat{Z}_{IS}}$ is

\begin{equation}
\log \widehat{Z}_{IS}(\bm{\theta}^{\ast})|\log \mathbf{\widehat{Z}_{IS}} \sim N( \bm{\theta}^{\ast}\bm{\beta}  + \mathbf{c}'\mathbf{C}^{-1}(\log\mathbf{\widehat{Z}_{IS}}-\bm{\psi}\bm{\beta} ), \sigma^{2}+\tau^2-\mathbf{c}'\mathbf{C}\mathbf{c}).
\label{condexp}
\end{equation}

\noindent Given true covariance parameters $( \sigma^{2},\phi, \tau^2 )$, a generalized least squares (GLS) estimator of regression parameter is $\widehat{\bm{\beta}}=(\bm{\psi}'\mathbf{C}^{-1}\bm{\psi})^{-1}\bm{\psi}'\mathbf{C}^{-1}\log\mathbf{\widehat{Z}_{IS}}$. By minimizing the mean square error, the best linear unbiased predictor (BLUP) for $\log\widehat{Z}_{IS}(\bm{\theta}^{\ast})$ can be derived as 

\begin{equation}
\log\widehat{Z}_{GP}(\bm{\theta}^{\ast}) =\bm{\theta}^{\ast}\widehat{\bm{\beta}} + \mathbf{c}'\mathbf{C}^{-1}(\log\mathbf{\widehat{Z}_{IS}} -\bm{\psi}\widehat{\bm{\beta}}),  
\label{BLUP1}
\end{equation}
where the mean squared error is 
\begin{equation}
\mathrm{Var}(\log\widehat{Z}_{GP}(\bm{\theta}^{\ast}) - \log\widehat{Z}_{IS}(\bm{\theta}^{\ast}) ) = \sigma^{2}+\tau^2 - \mathbf{c'Cc}+\mathbf{b}'(\bm{\psi}'\mathbf{C}^{-1}\bm{\psi})^{-1}\mathbf{b},~~~\mathbf{b}=\bm{\theta}^{\ast}-\bm{\psi}'\mathbf{C}^{-1}\mathbf{c}.
\label{BLUP2}
\end{equation}

\noindent Since covariance parameters $( \sigma^{2},\phi,\tau^2 )$ are unknown in practice, we can plug in estimates of these parameters (e.g. maximum likelihood or ordinary least squares) into the covariance $\mathbf{c},\mathbf{C}$, and GLS estimate $\widehat{\bm{\beta}}$. Using these plug-in estimates, \eqref{BLUP1} is called the empirical BLUP (EBLUP). With each iteration of the MCMC algorithm, this EBLUP is plugged into the log acceptance probability. The normalising function emulation algorithm is described in Algorithm~\ref{GPISalg}. 

\clearpage
\begin{algorithm}[hh]
\caption{normalising function emulation algorithm}\label{GPISalg}
\begin{algorithmic}
\normalsize

\State \textbf{Part 1: Construct two-stage approximation $\log \widehat{Z}_{GP}(\bm{\theta})$ to $\log Z(\bm{\theta})$ for any $\bm{\theta} \in \bm{\Theta}$}

\State {\it{Step 1.}} Construct $N$ MCMC algorithms, each with stationary distribution $h(\cdot|\bm{\widetilde{\theta}})/Z(\bm{\widetilde{\theta}})$. The last state of each of these Markov chains will be used in the step 2. 

\State {\it{Step 2.}} Calculate importance sampling approximation \eqref{IS} using the $N$ Markov chain samples for $\bm{\theta}^{(1)},\dots,\bm{\theta}^{(d)}$, to obtain, $\log\widehat{Z}_{IS}(\bm{\theta}^{(1)}),\dots,\log\widehat{Z}_{IS}(\bm{\theta}^{(d)})$.

\State {\it{Step 3.}} Obtain parameters  $( \sigma^2, \phi, \tau^2, \bm{\beta} )$ by fitting a Gaussian process via MLE to $(\bm{\theta}^{(1)}, \log\widehat{Z}_{IS}(\bm{\theta}^{(1)})), \dots,(\bm{\theta}^{(d)},\log\widehat{Z}_{IS}(\bm{\theta}^{(d)}))$.

\State \textbf{Part 2: MCMC algorithm with Gaussian process approximation.}

\State Given $\bm{\theta}_{n} \in \bm{\Theta}$ at $n$th iteration, construct the next step of the algorithm as follows 

\State {\it{Step 4.}} Propose $\bm{\theta}' \sim q(\cdot|\bm{\theta}_{n})$

\State {\it{Step 5.}} Evaluate $\log\widehat{Z}_{GP}(\bm{\theta}')$ from Gaussian process approximation as in \eqref{BLUP1} and accept $\bm{\theta}'$ with probability $\alpha$ where

$$\log\alpha = \min \left\lbrace \log \Big( \frac{p(\bm{\theta}')h(\mathbf{x}|\bm{\theta}')\widehat{Z}_{GP}(\bm{\theta})q(\bm{\theta}|\bm{\theta}')}{p(\bm{\theta})h(\mathbf{x}|\bm{\theta})\widehat{Z}_{GP}(\bm{\theta}')q(\bm{\theta}'|\bm{\theta})} \Big),0 \right\rbrace$$ else reject (set $\bm{\theta}_{n+1} = \bm{\theta}_n$). 
\end{algorithmic}
\end{algorithm}
\clearpage

This algorithm can dramatically reduce computing time because the two-stage approximations (Step 1 - 3 in Algorithm~\ref{GPISalg}) are precalculated and outside the MCMC algorithm. We can take advantage of parallel computation because constructing importance sampling estimates is embarrassingly parallel. Furthermore, the Gaussian process interpolation (Step 5 in Algorithm~\ref{GPISalg}) is extremely fast with each iteration of the MCMC algorithm. 

When the unnormalized likelihood $h(\mathbf{x}|\bm{\theta})$ is expensive to evaluate, it is computationally efficient to emulate the entire likelihood function instead of just the normalising function. We can construct log importance sampling estimates of likelihood functions for each particle as follows.  
 
\begin{equation}
\log \widehat{L}_{IS}(\bm{\theta}^{(i)}|\mathbf{x}) =
\log \frac{ h(\mathbf{x}|\bm{\theta}^{(i)})}{\widehat{Z}_{IS}(\bm{\theta}^{(i)})}
\label{LikIS}
\end{equation}
Then for a new $\bm{\theta^{\ast}}\in \Theta$, the log-likelihood value may be approximated in a similar fashion to \eqref{BLUP1}, resulting in $\log\widehat{L}_{GP}(\bm{\theta}^{\ast}|\mathbf{x})$. This approach can also be applied to the problem where likelihood evaluations are available but still expensive. In this case we do not need to construct importance  sampling estimates. The full likelihood emulation algorithm is described in Algorithm~\ref{GPISalg2}. 

\begin{algorithm}[hh]
\caption{Full likelihood function emulation algorithm}\label{GPISalg2}
\begin{algorithmic}
\normalsize

\State \textbf{Part 1: Construct two-stage approximation $\log\widehat{L}_{GP}(\bm{\theta}|\mathbf{x})$ to $
\log L(\bm{\theta}|\mathbf{x})$ for $\bm{\theta} \in \bm{\Theta}$}

\State {\it{Step 1.}} Construct $N$ independent MCMC algorithms, each with stationary distribution $h(\cdot|\bm{\widetilde{\theta}})/Z(\bm{\widetilde{\theta}})$. The last state of each of these Markov chains will be used in the step 2. 

\State {\it{Step 2.}} Calculate importance sampling approximation \eqref{LikIS} using the $N$ Markov chain samples for $\bm{\theta}^{(1)},\dots,\bm{\theta}^{(d)}$, to obtain, $ \log\widehat{L}_{IS}(\bm{\theta}^{(1)}|\mathbf{x}),\dots,\log\widehat{L}_{IS}(\bm{\theta}|\mathbf{x})$.

\State {\it{Step 3.}} Fit a Gaussian process to $(\bm{\theta}^{(1)}, \log\widehat{L}_{IS}(\bm{\theta}^{(1)}|\mathbf{x})), \dots,(\bm{\theta}^{(d)},\log\widehat{L}_{IS}(\bm{\theta}^{(d)}|\mathbf{x}))$ to obtain the parameters $( \sigma^2, \phi, \tau^2, \bm{\beta} )$. 

\State \textbf{Part 2: MCMC algorithm with Gaussian process approximation.}

\State Given $\bm{\theta}_{n} \in \bm{\Theta}$ at $n$th iteration, construct the next step of the algorithm as follows 

\State {\it{Step 4.}} Propose $\bm{\theta}' \sim q(\cdot|\bm{\theta}_{n})$

\State {\it{Step 5.}} Evaluate $\widehat{L}_{GP}(\bm{\theta}'|\mathbf{x})$ from Gaussian process approximation and accept $\bm{\theta}'$ with probability $\alpha$ where

$$\log\alpha = \min \left\lbrace \log \Big( \frac{p(\bm{\theta}')\widehat{L}_{GP}(\bm{\theta}'|\mathbf{x})q(\bm{\theta}|\bm{\theta}')}{p(\bm{\theta})\widehat{L}_{GP}(\bm{\theta}|\mathbf{x})q(\bm{\theta}'|\bm{\theta})} \Big),0 \right\rbrace$$ else reject (set $\bm{\theta}_{n+1} = \bm{\theta}_n$). 

\end{algorithmic}
\end{algorithm}

\subsection{Theoretical Justifications}

~~~~~For these function emulation approaches, we examine the  approximation error in terms of total variation distance \citep[cf.][]{mitrophanov2005sensitivity,alquier2014noisy}. Consider a target distribution $\pi(\bm{\theta}|\mathbf{x})$ whose Markov chain transition kernel is $\mathbf{P}$. By plugging in $\log\widehat{Z}_{IS}(\bm{\theta})$ into the log acceptance probability, the first-stage approximated transition kernel $\mathbf{\widehat{P}}_{IS}$ can be constructed, the stationary distribution of which is $\widehat{\pi}_{IS}(\bm{\theta}|\mathbf{x})$. The second-stage approximated kernel $\mathbf{\widehat{P}}_{GP}$ is constructed by replacing the $\log\widehat{Z}_{IS}(\bm{\theta})$ with $\log\widehat{Z}_{GP}(\bm{\theta})$ and $\widehat{\pi}_{GP}(\bm{\theta}|\mathbf{x})$ is the corresponding stationary distribution. We make the following assumptions.

\begin{assumption}
\label{assumption1}
$\exists$ constant $c_{p}>1$ s.t. $1/c_{p} \leq p(\bm{\theta}) \leq c_{p}$.
\end{assumption}
\begin{assumption}
\label{assumption2}
$\exists$ constant $c_{q}>1$ s.t. $1/c_{q} \leq q(\bm{\theta}'|\bm{\theta}) \leq c_{q}$.
\end{assumption}
\begin{assumption}
\label{assumption3}
$\exists$ constant $k$,$K$ s.t. $k \leq h(\mathbf{x}|\bm{\theta}) \leq K$.
\end{assumption}
\begin{assumption}
\label{assumption4}
$\bm{\Theta}$ is compact.
\end{assumption}

In many applications, and this is the case for the examples discussed in Section 4, the sample space $\mathcal{X}$ may be reasonably assumed to be finite and the parameter space $\bm{\Theta}$ may be assumed to be a compact set (assumption 4). Hence, the assumptions 1-3 may also be easily checked. Theorem~\ref{GPproof} quantifies the total variation distance between the target posterior distribution and the two-stage approximated Markov transition kernel. 

\begin{theorem}
\label{GPproof}
Consider Markov transition kernel $\mathbf{\widehat{P}}_{GP}$ constructed by plugging in $\log \widehat{Z}_{GP}(\bm{\theta})$ into the log acceptance probability. Suppose 
Assumptions \ref{assumption1} to \ref{assumption4} hold. Then $\|\pi(\cdot|\mathbf{x})-\delta_{\bm{\theta}_{0}}\mathbf{\widehat{P}}_{GP}^{n}\|_{TV} \leq \rho^{n}M + \epsilon(N) + \epsilon(d)$ almost surely for bounded constant $M$ and $0<\rho<1$.
\end{theorem}

Proof of Theorem~\ref{GPproof} is provided in the supplementary material. Given the result in this theorem, the Markov chain samples from the normalising function emulation algorithm will be close to the target distribution $\pi(\bm{\theta}|\mathbf{x})$, as the sample size for importance sampling estimates ($N$) and the number of particles ($d$) are increased  ($\epsilon(N)$ and $\epsilon(d)$ goes to 0 as $N$ and $d$ increases respectively). We note that learning  how $\epsilon$ scales with $N$ and $d$ is also potentially of interest, but this is quite problem-specific and poses challenges. For fixed $N$ and $d$, the stationary distribution of the proposed algorithm is $\mathbf{\widehat{\pi}}_{GP}(\bm{\theta}|\mathbf{x})$, which is different from  the desired target $\pi(\bm{\theta}|\mathbf{x})$. Hence, in practice, this algorithm is asymptotically inexact. However, with an  appropriate choice of $N$ and $d$ this algorithm appears to provide reasonable approximations more quickly than other asymptotically inexact algorithms, as is evident from numerous applications in Section 4.

In similar fashion, we examine the approximation error for Algorithm 2, the likelihood function emulation approach, in Corollary~\ref{GPproof2}. With increasing number of $N$ and $d$, the posterior recovered from a likelihood function emulation approach becomes close to the true target distribution $\pi(\bm{\theta}|\mathbf{x})$. For finite $N$ and $d$, this algorithm is also asymptotically inexact. Proof of Corollary~\ref{GPproof2} is in the supplementary material.
  
\begin{corollary}
\label{GPproof2}
Consider Markov transition kernel $\mathbf{\widehat{P}}_{GP}$ constructed by plugging in $\log \widehat{L}_{GP}(\bm{\theta}|\mathbf{x})$ into the log acceptance probability. Suppose that
Assumptions \ref{assumption1} to \ref{assumption4} hold. Then $\|\pi(\cdot|\mathbf{x})-\delta_{\bm{\theta}_{0}}\mathbf{\widehat{P}}_{GP}^{n}\|_{TV} \leq \rho^{n}M + \epsilon(N) + \epsilon(d)$ almost surely for bounded constant $M$ and $0<\rho<1$.
\end{corollary}

\subsection{Pre-MCMC Details for the  Function Emulation Algorithms}

~~~~~The preliminary non-MCMC part of the function emulation algorithms involve constructing the two-stage approximation. For this, the particles $\bm{\psi}=( \bm{\theta^{(1)}},\dots, \bm{\theta^{(d)}} )$ should be chosen so that they cover well the important regions of the parameter space $\bm{\Theta}$. The choice of particles is important for both algorithms. In general, we found that a short run of the double Metropolis-Hastings (DMH) \citep{liang2010double} was useful in providing particles. This was an approach also used in \cite{liang2015adaptive}. The DMH Algorithm may be described as follows:

\begin{algorithm}
\caption{Double Metropolis-Hastings algorithm for choosing particles}\label{DMHalg}
\begin{algorithmic}
\normalsize
\State Given $\bm{\theta}_{n} \in \bm{\Theta}$ at $n$th iteration.

\State {\it{Step 1.}} Propose $\bm{\theta}' \sim q(\cdot|\bm{\theta}_{n})$.

\State {\it{Step 2.}} Generate the auxiliary variable approximately from probability model at $\bm{\theta}'$: $\mathbf{y} \sim h(\cdot|\bm{\theta}')/Z(\bm{\theta}')$ using Metropolis-Hastings updates.

\State {\it{Step 3.}} Accept $\bm{\theta}_{n+1}=\bm{\theta}'$ with probability

$$\alpha = \min\left\lbrace \frac{p(\bm{\theta}')h(\mathbf{x}|\bm{\theta}')h(\mathbf{y}|\bm{\theta}_{n})q(\bm{\theta}_{n}|\bm{\theta}')}{p(\bm{\theta}_{n})h(\mathbf{x}|\bm{\theta}_{n})h(\mathbf{y}|\bm{\theta}')q(\bm{\theta}'|\bm{\theta}_{n})}, 1 \right\rbrace$$ else reject (set $\bm{\theta}_{n+1}=\bm{\theta}_n$).

\State Repeat Step 1 - Step 3 until we have $d$ accepted samples $\lbrace \bm{\theta}^{(1)},\dots,\bm{\theta}^{(d)}\rbrace$.

\end{algorithmic}
\end{algorithm}

There are other approaches to choosing particles. When the summary statistics are low-dimensional $S(\mathbf{x})$ (e.g. the exponential random graph models), we recommend the approximate Bayesian computation (ABC) algorithm \citep{beaumont2002approximate} as in \cite{jin2013bayesian}. Starting from a wide domain $\mathcal{D}_{1}$ Algorithm~\ref{ABCalg} can search interesting region of parameter space $\mathcal{D}_{2}$. Then $d$ number of particles are generated over $\mathcal{D}_{2}$. In Step 3 of the Algorithm~\ref{ABCalg}, auxiliary variables can be generated via parallel computation. In Step 4 $\epsilon$ is a tolerance, which controls the trade-off between computational efficiency and accuracy. With decreasing $\epsilon$, we can have ABC samples from the distribution which is close to the posterior distribution, but acceptances will be rare. We set tolerance $\epsilon$ equal to about the 0.03 quantile of the Euclidean distance between simulated and observed summary statistics $(\|S(\mathbf{y}^{(i)})-S(\mathbf{x})\|)$ in our study.

\begin{algorithm}
\caption{Approximate Bayesian Computation for choosing particles }\label{ABCalg}
\begin{algorithmic}
\normalsize
\State {\it{Step 1.}} Select a wide rectangular domain $\mathcal{D}_{1}$ over the parameter space using the MPLE \citep{besag1974spatial} and its standard error: $\mathcal{D}_{1}=[\widehat{\bm{\theta}}_{MPLE} - 10\widehat{\bm{\sigma}}, \widehat{\bm{\theta}}_{MPLE} + 10\widehat{\bm{\sigma}}]$.

\State {\it{Step 2.}} Generate $D$ Latin hypercube design points $\lbrace \bm{\nu}^{(1)},..., \bm{\nu}^{(D)}\rbrace$ over $\mathcal{D}_{1}$.

\State {\it{Step 3.}} Simulate the auxiliary variable for each $\bm{\nu}^{(i)}$: 

\noindent $\mathbf{y}^{(i)} \sim h(\cdot|\bm{\nu}^{(i)})/Z(\bm{\nu}^{(i)})$ using Metropolis-Hastings updates for $i=1,...,D$.

\State {\it{Step 4.}} Without loss of generality, choose $\lbrace \bm{\nu}^{(1)},..., \bm{\nu}^{(d_{1})}\rbrace$ from $\lbrace \bm{\nu}^{(1)},..., \bm{\nu}^{(D)}\rbrace$ which satisfy $\|S(\mathbf{y}^{(i)})-S(\mathbf{x})\| < \epsilon$, where $\| \cdot \|$ denotes the Euclidean distance. 

\State {\it{Step 5.}} Select a smaller rectangular domain $\mathcal{D}_{2} \subset  \mathcal{D}_{1}$ which cover the important region of the parameter space: $\mathcal{D}_{2}=[\min_{j=1,\dots,d_{1}}{\lbrace \bm{\nu}^{(j)}\rbrace},\max_{j=1,\dots,d_{1}}{\lbrace \bm{\nu}^{(j)}\rbrace}]$.

\State {\it{Step 6.}} Generate $d$ number of particles $\lbrace \bm{\theta}^{(1)},..., \bm{\theta}^{(d)}\rbrace$ over $\mathcal{D}_{2}$ using Latin hypercube design.

\end{algorithmic}
\end{algorithm}

If we use Algorithm~\ref{DMHalg} or Algorithm~\ref{ABCalg} to generate particles, $d = 400$ particles seems to work well in practice for problems with up to 4-dimensional parameter spaces. Then the number of samples $(N)$ for constructing importance sampling estimate should be specified. Considering that our approach is asymptotically inexact, a conservative approach involves using a large value of $N$. In this manuscript we set $N=1,000$ to $2,000$.

\section{Applications}

~~~~~We apply our approach to two general classes of models with intractable normalising functions: (1) an exponential random graph model, and (2) an attraction-repulsion point process model. Double Metropolis Hastings (DMH) was  found to be the most efficient among current algorithms in terms of effective sample size per time. Hence, to illustrate the computational and statistical efficiency of our approach, we compare the normalising function emulation (NormEm) and likelihood function emulation (LikEm) algorithms with DMH. For a  large social network example, DMH is too expensive to be practical, but both NormEm and LikEm take under 2 hours, a dramatic computational gain.

Our function emulation approach (LikEm) is more broadly applicable than to just doubly intractable distributions. To illustrate this, we apply this method to a susceptible-infected-recovered infectious disease model where likelihood evaluations are available but computationally expensive. The different examples we study illustrate different computational challenges. The code for our algorithms is implemented in {\tt R} \citep{ihaka1996r} and {\tt C++}, using the \texttt{Rcpp} and \texttt{RcppArmadillo} packages \citep{eddelbuettel2011rcpp}. We fit Gaussian process models to estimate hyper-parameters $( \sigma^2, \phi, \tau^2, \bm{\beta} )$ using the \texttt{DiceKriging} package \citep{roustant2012dicekriging}. The point estimates in each example are simply means of the entire sample; there was no thinning or burn-in. The highest posterior density (HPD) is calculated by using \texttt{coda} package in {\tt R}. The calculation of Effective Sample Size (ESS) follows \cite{kass1998markov,robert2013monte}. For a point process model and an infectious disease model, we calculate the total variational (TV) distance of the marginal posterior distributions between each of the algorithms and a gold standard using \texttt{density} function in {\tt R}. We cannot construct a gold standard for a social network model because of the computational expense. Figures for bivariate and univariate posterior densities are in the supplementary material. All the code was run on dual 10 core Xeon E5-2680 processors on the Penn State high performance computing cluster. The source code may be downloaded from the following repository (https://github.com/jwpark88/FuncEmul).

\subsection{Social Network Models}
~~~~~Exponential random graph models (ERGM) \citep{robins2007introduction, hunter2008ergm} describe relationships among actors in networks. Consider the undirected ERGM with $n$ nodes. For all $i \neq j$, $x_{i,j}=1$ if the $i$th node and $j$th node are connected, otherwise $x_{i,j}=0$ and $x_{i,i}$ is defined as 0. Calculation of the normalising function requires summation over all $2^{n(n-1)/2}$ network configurations, which is intractable. Consider the ERGM, where the probability model is

\begin{equation}
L(\bm{\theta}|\mathbf{x})=\frac{1}{Z(\bm{\theta})}\exp\left\lbrace \theta_{1}S_{1}(\mathbf{x}) + \theta_{2}S_{2}(\mathbf{x}) \right\rbrace,
\label{fauxmagnolia}
\end{equation}

\begin{gather*}
S_{1}(\mathbf{x})=\sum_{i=1}^{n}{x_{i+} \choose 1} ~~~~ S_{2}(\mathbf{x})=e^{\tau}\sum_{k=1}^{n-2}\left\lbrace 1-(1-e^{-\tau})^{k} \right\rbrace ESP_{k}(\mathbf{x})
\end{gather*}

\noindent where $S_{1}(\mathbf{x})$ is the number of edges and $S_{2}(\mathbf{x})$ is the geometrically weighted edge-wise shared partnership (GWESP) statistic \citep{hunter2006inference,hunter2007curved}. $ESP_{k}(\mathbf{x})$ denotes the number of connected pairs $(i,j)$, where $i$ and $j$ have $k$ common neighbors. $ESP_{k}(\mathbf{x})$ models high-order transitivites, because $ESP_{k}(\mathbf{x})$ is a function of triangles. Therefore, GWESP models edge-wise shared partnership by placing geometric weights $\tau=0.25$ on the edges with higher transitivites. We used uniform prior distribution for $\theta_{1}$ and $\theta_{2}$ with ranges of $[-7.8,-6.8]$ and $[1.8,2.5]$ respectively. These priors were centered around the MPLE with a width of 10 standard deviations. For this model we can generate auxiliary variables (DMH) or Monte Carlo samples for importance sampling estimates (NormEm and LikEm) via Gibbs updates. For each iteration, $(i,j)$ pairs are chosen randomly and $x_{i,j}$ is set to 0 or 1 according to the full conditional probabilities. See \cite{hunter2008ergm} for details. Here we study this model for both real and simulated data examples. 

We study the Faux Magnolia high school data set \citep{resnick1997protecting}, which describes an in-school friendship network among 1461 students. For all the algorithms, samples are generated from the probability model through 1 cycle of Gibbs updates. In both function emulation approaches, particles are selected via approximate Bayesian computation (ABC) as in Algorithm~\ref{ABCalg}. We initially generate $D=3,000$ Latin hypercube design points over $\mathcal{D}_{1}$. We set a tolerance $\epsilon$ equal to 0.03 quantile to obtain $\mathcal{D}_{2}$, the important region of the parameter space. Then we generate $d=400$ particles over $\mathcal{D}_{2}$ by using a Latin hypercube design. Then $N=1,000$ numbers of samples are used to construct importance sampling estimates. We used parallel computing to obtain importance sampling estimates. The parallel computing was implemented through {\tt OpenMP }\citep{dagum1998openmp} with the samples generated in parallel across 20 processors. For all the algorithms we use a multivariate normal proposal. The covariance used in the multivariate normal is obtained as follows. The initial covariance matrix for the proposal is the inverse of the negative hessian matrix from the MPLE. We re-estimate the sample covariance matrix for the first 10,000 iterations and use this in the proposal. After 10,000 iterations the covariance of the proposal is fixed for the rest of the algorithm. Algorithms were run until the Monte Carlo standard errors  calculated by batch means \citep{jones2006fixed,flegal2008markov} are below 0.001. 

\begin{table}[hh]
\centering
\begin{tabular}{ccc}
  \hline
DMH & $\theta_{1}$ & $\theta_{2}$ \\
  \hline
Mean & -7.47 & 2.31   \\
95\%HPD &  (-7.56, -7.38) &  ( 2.21 ,2.41)   \\
ESS &  1568.90 & 1470.91\\
Time(hour) & 41.41 \\
minESS/Time & 35.52\\
  \hline
NormEm & $\theta_{1}$ & $\theta_{2}$ \\
  \hline
Mean &  -7.47 & 2.31 \\
95\%HPD &  (-7.55, -7.38) & (2.21,2.41)  \\
ESS &  2509.00 & 2332.76\\
Time(hour) & 1.07\\
minESS/Time &  2163.92\\
  \hline
LikEm & $\theta_{1}$ & $\theta_{2}$ \\
  \hline
Mean &   -7.47 & 2.31\\
95\%HPD &  (-7.55, -7.38) & (2.21,2.41)  \\
ESS &  2460.23 &   2552.40\\
Time(hour) & 0.90\\
minESS/Time &  2738.60\\
  \hline  
\end{tabular}
\caption{Results for parameter estimates in ERGM for a Faux Magnolia high school data set. 25,000 MCMC samples are generated from each algorithm.}
\label{ergmout}
\end{table}

The  function emulation approaches dramatically reduce computational time even when compared to the fastest algorithm, DMH. DMH takes about 40 hours but both function emulation approaches only take about 1 hour to run, including pre-computing time. Table~\ref{ergmout} indicates that the estimates from the different algorithms are similar. Bivariate and univariate posterior densities are illustrated in the supplementary material. We can also account for mixing of the algorithms  through effective sample size (ESS),  \citep{kass1998markov} which approximates the number of independent samples that correspond to the number of dependent samples from the chain (a chain with very low dependence would return an ESS very similar to the actual Markov chain length). When accounting for mixing time, as shown in Table~\ref{ergmout}, the proposed algorithms show larger ESS than DMH for the same length. Naturally, the differences in minimum effective sample size per time (minESS/T) are even more dramatic. In summary, the function emulation approaches are much faster than current algorithms, result in better mixing  chains, and provide reasonable results. 

To validate our methods, we simulated a $2,000$ by $2,000$ network via 100 cycles of Gibbs updates, where the true parameter is $( \theta_{1}, \theta_{2} ) = ( -7, 2 )$. We study our methods for different combinations of $N$ and $d$. The rest of the settings for all algorithms are identical to the real data example. Here we only provide the results for $\theta_{2}$ because similar results are observed for the other parameter. We provide results for $\theta_{1}$ in the supplementary material. In Table~\ref{ergmoutsimul} we observe that both function emulation approaches can recover the true parameter value used in the simulation, across different choices of $N$ and $d$. Implementing DMH is infeasible because auxiliary variable simulations are computationally expensive for this example. Based on our preliminary run, we estimate that it will take at least 5 days to run. Considering that DMH is the fastest approach among existing algorithms \citep{park2018bayesian}, this highlights the fact that our approach can provide reliable results for large networks, and do so much faster than current approaches.

\begin{table}[hh]
\centering
\begin{tabular}{cccccccccc}
  \hline
$\theta_{2}$ & $N$ & $d$ & Mean & 95\%HPD & ESS & Time(hour) & ESS/Time\\
  \hline
DMH & NA& NA & NA & NA & NA & NA &  NA\\
NormEm & 1000 & 400 & 2.01 & (1.96, 2.06) & 2248.88 & 1.54 & 1459.39 \\
& 1000 & 200 & 2.01 & (1.96, 2.06) & 2447.13 & 1.54 & 1588.80\\
& 1000 & 100 & 2.01 & (1.96, 2.06) & 2459.70 & 1.54 & 1597.34\\
 & 500 & 400 & 2.01 & (1.96, 2.06) & 2372.00 & 1.35 &  1755.24\\
 & 500 & 200 & 2.01 & (1.96, 2.06) & 2371.55 & 1.35 &  1756.02\\
 & 500 & 100 & 2.01 & (1.96, 2.07) & 2489.17 & 1.35 & 1821.87 \\
LikEm  & 1000 & 400 & 2.01 & (1.96, 2.06) & 2570.34 & 1.67 & 1535.80 \\
 & 1000 & 200 & 2.01 & (1.96, 2.06) & 2411.25 & 1.67 & 1441.29\\
 & 1000 & 100 & 2.01 & (1.96, 2.06) & 2552.63 & 1.67 &  1526.16\\
 & 500 & 400 & 2.01 & (1.96, 2.06) & 2604.7 & 1.45 & 1793.38\\
 & 500 & 200 & 2.01 & (1.96, 2.06) & 2374.33 & 1.45 & 1635.83\\
 & 500 & 100 & 2.01 & (1.96, 2.06) & 2546.33 & 1.45 & 1759.24\\
   \hline
Simulated Truth & 2.00\\
   \hline
\end{tabular}
\caption{Results for $\theta_{2}$ in ERGM for a large simulated network. 25,000 MCMC samples are generated from each algorithm. The true $\theta_2$ value is 2. DMH was too expensive to be practical.}
\label{ergmoutsimul}
\end{table}
\clearpage

\subsection{An Attraction-Repulsion Point Process Model}

~~~~~A spatial point process in two dimensions is a random set of points in a bounded plane $S \subset R^{2}$. Consider a realisation of points $\mathbf{x}=( x_{1},\dots,x_{n} )$, and $D_{ij}$ is the distance between the coordinates of $x_{i}$ and $x_{j}$. Then a probability model can describe spatial patterns among the points by introducing an interaction function $\phi(D_{ij})$. \cite{goldstein2014attraction} extends the Strauss process  \citep{strauss1975model} to develop a model describing both attraction and repulsion patterns of the cells infected with human respiratory syncytial virus (RSV). The interaction function is

\begin{equation}
\phi(D) = \begin{cases}
      0 & 0 \leq D \leq R \\
      \theta_{1}-\left(\frac{\sqrt{\theta_{1}}}{\theta_{2}-R}(D-\theta_{2}) \right)^{2}  & R< D \leq D_{1} \\
      1+\frac{1}{(\theta_{3}(D-D_{2}))^{2}} & D > D_{1}
\end{cases}
\label{e26}
\end{equation}
and the probability model is 
\begin{equation}
L(\bm{\theta}|\mathbf{x})=\frac{\lambda^{n} \left[\prod_{i=1}^{n} \exp\left\lbrace \min\left(\sum_{i\neq j}\log{(\phi(D_{i,j}))},1.2\right)  \right\rbrace \right]}{Z(\bm{\theta})}, ~~ \bm{\theta}=( \lambda,\theta_{1},\theta_{2},\theta_{3} ).
\label{e27}
\end{equation}

\noindent The intensity of the point process is controlled by $\lambda$, and $( \theta_{1},\theta_{2},\theta_{3} )$ control the interaction function. $\theta_{1}$ is the peak value of $\phi$, $\theta_{2}$ is value of $D$ at the peak of $\phi$ and $\theta_{3}$ represents the descent rate after the peak. We used uniform priors with range $[2\times10^{-4},5\times 10^{-4}]\times[1,2]\times[10,20]\times[0,1]$ for $(\lambda, \theta_{1},\theta_{2},\theta_{3})$, which is a plausible range obtained from \cite{goldstein2014attraction}.
The model is able to explain both attraction and repulsion spatial associations among infected cells. Calculation of the normalising function requires integration over the continuous domain $S$, which is infeasible. For this model we can generate auxiliary variables (DMH) or Monte Carlo samples for importance sampling estimates (NormEm and LikEm) via birth-death MCMC \citep{geyer1994simulation}. At each iteration of the chain, an existing point is removed (death) or a new point is added (birth) with equal probability. See \cite{goldstein2014attraction} for details. We study inference for this model in the context of real and simulated data examples.

We study the RSV A point pattern ($n \approx 3,000$) from $\mathbf{1A2A}$ experiment (RSV A primary virus, RSV A secondary virus), with a 16 hour time lag \citep{goldstein2014attraction}.  For all the algorithms, samples are generated from the probability model through 10 cycles of birth-death MCMC. For both function emulation approaches, $N = 2,000$ samples are used to construct importance sampling estimates and $d = 400$ particles are used for Gaussian process approximations. Importance sampling estimates are obtained in parallel as in the previous example. We run the DMH algorithm (Algorithm~\ref{DMHalg}) until we obtain $d=400$ unique particles; this took 2,412 iterations with the first 1,000 iterations are discarded for burn-in. For all the algorithms we use a multivariate normal proposal with covariance matrix obtained as in the social network example. Although the DMH algorithm is asymptotically inexact, by increasing the number of iterations for the birth-death MCMC, the algorithm can provide more accurate results \citep{liang2015adaptive}. Since other asymptotically exact algorithms are infeasible  for this example as pointed out in \cite{park2018bayesian}, we treated a run from DMH as our gold standard; it was run for 100,000 iterations with 20 cycles of birth-death MCMC. All algorithms were run until the Monte Carlo standard error is at or below 0.001.

\clearpage
\begin{table}[hh]
\centering
\begin{tabular}{ccccc}
  \hline
DMH & $\lambda \times 10^4$ & $\theta_{1}$ & $\theta_{2}$ & $\theta_{3}$\\
  \hline
Mean & 2.97 & 1.34 & 11.52 & 0.22 \\
95\%HPD & (2.64, 3.30) & (1.30, 1.39) & (10.68, 12.28) & (0.17, 0.28) \\
TV &  0.23 & 0.37 & 0.28 & 0.23 \\
ESS &  977.43 & 1037.47 & 1120.68 & 899.01\\
Time(hour) & 18.99\\
minESS/Time & 47.35\\
  \hline
NormEm & $\lambda \times 10^4$ & $\theta_{1}$ & $\theta_{2}$ & $\theta_{3}$ \\
  \hline
Mean &   2.96 & 1.34 & 11.51 & 0.22 \\
95\%HPD & (2.62, 3.27) & (1.29, 1.39) & (10.65, 12.33) & (0.17, 0.27)\\
TV & 0.26 & 0.28 & 0.33 & 0.24 \\
ESS & 1996.07 & 2151.35 & 2167.10 & 1564.01 \\
Time(hour) & 3.80\\
minESS/Time & 411.10 \\
  \hline
LikEm & $\lambda \times 10^4$ & $\theta_{1}$ & $\theta_{2}$ & $\theta_{3}$  \\
  \hline
Mean & 2.98 & 1.34 & 11.50 & 0.22 \\
95\%HPD & (2.61, 3.35) & (1.30, 1.39) & (10.65, 12.31) & (0.17, 0.29)\\
TV &  0.47 & 0.37 & 0.30 & 0.40 \\
ESS & 1883.62 &  1988.59 & 1815.97 &  1315.89 \\
Time(hour) & 2.52\\
minESS/Time &  521.46 \\
  \hline  
Gold & $\lambda \times 10^4$ & $\theta_{1}$ & $\theta_{2}$ & $\theta_{3}$  \\
  \hline
Mean &  2.97 & 1.34 & 11.49 & 0.22  \\
95\%HPD & (2.64, 3.28) & (1.30, 1.39) & (10.68, 12.27) & (0.17, 0.28)\\
\hline
\end{tabular}
\caption{Inference results for the RSV point pattern from $\mathbf{1A2A}$ experiment. 40,000 MCMC samples are generated from each algorithm.}
\label{attractionout} 
\end{table}
\clearpage

While  DMH takes about 19 hours, the function emulation approaches take between 2.5 and 4 hours. LikEm is about an hour faster than NormEm. This is because the unnormalized likelihood is expensive to evaluate in a point process example. For the ERGM, we can evaluate $h(\mathbf{x}|\bm{\theta})$ simply by taking the product of $\bm{\theta}$ and $S(\mathbf{x})$ once we evaluate $S(\mathbf{x})$. However for the attraction repulsion point process model, $\phi(\cdot)$ needs to be reevaluated at the distance matrix of $\mathbf{x}$ with different parameters to calculate $h(\mathbf{x}|\bm{\theta})$. Here, even though this is a normalising function problem, emulating the entire likelihood function is  helpful. For a 2-dimensional ERGM example, our approach is about 40 times faster than DMH, while in the 4-dimensional attraction-repulsion point process problem, our approach is about 6-7 times faster than DMH. This difference comes from the precomputation step. Obtaining particles takes much more time for the point process example. Because a point process model neither has low-dimensional summary statistics nor analytical gradients, a short run of DMH needs to be used to obtain particles, which takes 1.5 hours. Table~\ref{attractionout} indicates that the estimates from the algorithms are well matched to the gold standard. Bivariate and univariate posterior densities are illustrated in the supplementary material. To measure the accuracy of our algorithms we calculate the total variational (TV) distance of the marginal posterior distributions between each of the algorithms and a gold standard. TVs for function emulation approaches are comparable with those of DMH. Compared to DMH, function emulation approaches can achieve the same accuracy within a much shorter time. In Table~\ref{attractionout}, the function emulation approaches show larger ESS than DMH. When accounting for mixing, the difference increases, as is apparent from a comparison of minimum effective sample size per second (minESS/T). In  summary, our algorithm is much faster and provides reasonable inference results. LikEm, in particular, has significant computational advantages over DHM.

To validate our methods, a point pattern ($n\approx 3,000$) is simulated under RSV B settings in \cite{goldstein2014attraction}. A point process is simulated by 100 cycles of birth-death MCMC, where the true parameter is $( \lambda \times 10^4, \theta_{1}, \theta_{2}, \theta_{3} ) = ( 4, 1.2, 15, 0.3 )$. We study our approaches for different combinations of $N$ and $d$. The rest of the settings for all algorithms are implemented using the same tuning parameters as in the real data example. Here we only provide the inference results regarding $\theta_{3}$ because similar results are observed for the other parameters. We provide results for other parameters in the supplementary material. Table~\ref{attractionoutsimul} indicates that the estimates from the function emulation approaches are similar to the those of the gold standard, when we have large enough $N$ and $d$ ($N=2,000$ and $d=200,400$ or $N=1,000$ and $d=400$). We observe that TVs for function emulation approaches are comparable with those of DMH for these $N$ and $d$ values. Otherwise, recovered posteriors from the function emulation approaches do not match the gold standard (TVs $>1$). This fact demonstrates that with increasing parameter dimensions, function emulation approaches become more sensitive to the choice of $N$ and $d$. Considering that emulation approaches are much cheaper than DMH, for parameter dimensions of 4 and under, we recommend using $N=2,000$ and $d=400$. 

\clearpage
\begin{table}[hh]
\centering
\begin{tabular}{cccccccccc}
  \hline
$\theta_{3}$ & N & d & Mean & 95\%HPD & TV & ESS  & Time(hour) & ESS/Time\\
  \hline
DMH & NA & NA & 0.34 & (0.21, 0.49) & 0.18 & 784.58 & 22.10 & 35.50\\
NormEm & 2,000 & 400 & 0.34 & (0.20, 0.49) & 0.25 &  931.96 & 4.10 & 227.30\\
 & 2,000 & 200 & 0.34 & (0.20, 0.48) & 0.27 & 1250.63 & 3.18 &  393.28\\
 & 2,000 & 100 & 0.21 & (0.20, 0.21) & 2.42 & 469.78 & 2.18 & 215.50 \\
 & 1,000 & 400 & 0.33 & (0.20, 0.48) & 0.37 & 973.40 & 3.49 & 278.91 \\
 & 1,000 & 200 & 0.26 & (0.20, 0.44) & 2.89 & 645.31 & 2.25 & 286.81\\
 & 1,000 & 100 & 0.20 & (0.20, 0.20) & 2.50 & 531.26 & 1.63 & 325.93\\
  LikEm  & 2,000 & 400 & 0.34 & (0.21, 0.48) & 0.30 & 1358.18 & 2.87 & 473.90\\ 
 & 2,000 & 200 & 0.33 & (0.21, 0.49) & 0.71 & 1274.48 & 2.04 & 624.75\\
 & 2,000 & 100 & 0.45 & (0.21, 0.85) & 3.49 & 1494.04 & 1.62 & 922.25\\
 & 1,000 & 400 & 0.34 & (0.21, 0.49) & 0.35 & 1116.76 & 2.26 & 494.14 \\
 & 1,000 & 200 & 0.51 & (0.21, 0.91) & 5.79 & 1748.24 & 1.43 & 1222.54\\
 & 1,000 & 100 & 0.59 & (0.20, 0.96) & 5.79 & 1677.07 & 1.02 & 1644.18 \\
 Gold & NA & NA & 0.34 & (0.21, 0.50) \\
   \hline
Simulated Truth & 0.30 \\
   \hline
\end{tabular}
\caption{Inference results for outputs for $\theta_{3}$ in simulated attraction repulsion point process model. 40,000 MCMC samples are generated from each algorithm.}
\label{attractionoutsimul} 
\end{table}

\subsection{Susceptible-Infected-Recovered Models}

~~~~~Our function emulation approach may be more broadly applicable to any likelihood function that is expensive to evaluate. Susceptible-infected-recovered (SIR) compartmental models \citep{dietz1967epidemics} are widely used to quantify the dynamics of infectious diseases. \cite{park2017ensemble} examine the rotavirus disease for children under 5 years of age in Niger with several variants of SIR compartmental models. For these models, the evaluation of the likelihood is available but computationally expensive. This is because for each set of parameter values, the periodic solution to the SIR dynamic equation is required for the likelihood calculation.  Our study shows that LikEm provides comparable results, and is 10 times faster than the regular MCMC algorithm. We provide details in the the supplementary material.

\subsection{Computational Complexity}

~~~~~We examine the computational complexity of the function emulation approaches (NormEm and LikEm) and double Metropolis-Hastings (DMH) in ERGM and interaction point process models, summarizing how our algorithms scale with an increase in the size of the data, $n$. We denote by $n$ the number of nodes for ERGM and the number of points for the attraction-repulsion point process. 

\begin{table}[tt]
\centering
\begin{tabular}{ccc}
  \hline
 &  Simulated Truth & Network Density$\times 10^3$\\
  \hline
200 nodes &  $(-6.0, 2.0)$ & 4.12\\
282 nodes &  $(-6.1, 2.0)$ & 3.71 \\
400 nodes &  $(-6.2, 2.0)$ & 4.45 \\
565 nodes &  $(-6.3, 2.0)$ & 4.23 \\\
800 nodes &  $(-6.4, 2.0)$ & 4.54 \\
   \hline
\end{tabular}
\caption{Simulation settings with different scales of networks.}
\label{ERGMsetting} 
\end{table}

We begin with a few caveats. The computational complexity of the ERGM used in the manuscript not only is dependent on the number of nodes but also is dependent on the density of the network. The density of a network is defined as $S_{1}(\mathbf{x})/{n \choose 2}$ where $S_{1}(\mathbf{x})$ is the number of edges. Therefore, true parameter values are selected to maintain similar density of a network with different scales of simulated networks as in Table~\ref{ERGMsetting}. To simplify calculations we assume the dimensions of the data and the simulated data are the same. We note that while this always holds for ERGMs, in the interaction point process case this is not always true as the simulated data is generated through birth-death MCMC with varying dimensions.  

The main difference in calculating complexity of both examples comes from the different structure of $h(\mathbf{x}|\bm{\theta})$. For ERGM, we can take the product of $\bm{\theta}$ and summary statistics for evaluation of the unnormalized likelihood function in different $\bm{\theta}$. In the point process example, however, $h(\mathbf{x}|\bm{\theta})$ requires recalculation of the interaction function with different parameters. Here we provide our main observations (see supplement for details), where costs are per iteration of the main MCMC algorithm: (1) In the ERGM, complexity of DMH is $\mathcal{O}(n^3)$. Ignoring pre-MCMC (particle finding and importance sampling) costs, complexity for both function emulation approaches is $\mathcal{O}(d^3)$, where $d$ is the number of particles. (2) Complexity for the point process model is $\mathcal{O}(n^2)$ for DMH and NormEm. However the amount of calculations per iteration of NormEm is about 1/5 that of DMH. Ignoring pre-MCMC costs, the complexity of LikEm is $\mathcal{O}(d^3)$, because we can avoid expensive $h(\mathbf{x}|\bm{\theta})$ evaluation in the MCMC steps. (3) Pre-MCMC costs are heavily parallelizable in both problems, and become marginal with increasing number of available cores. 

\begin{figure}
\begin{center}
\includegraphics[ scale = 0.6]{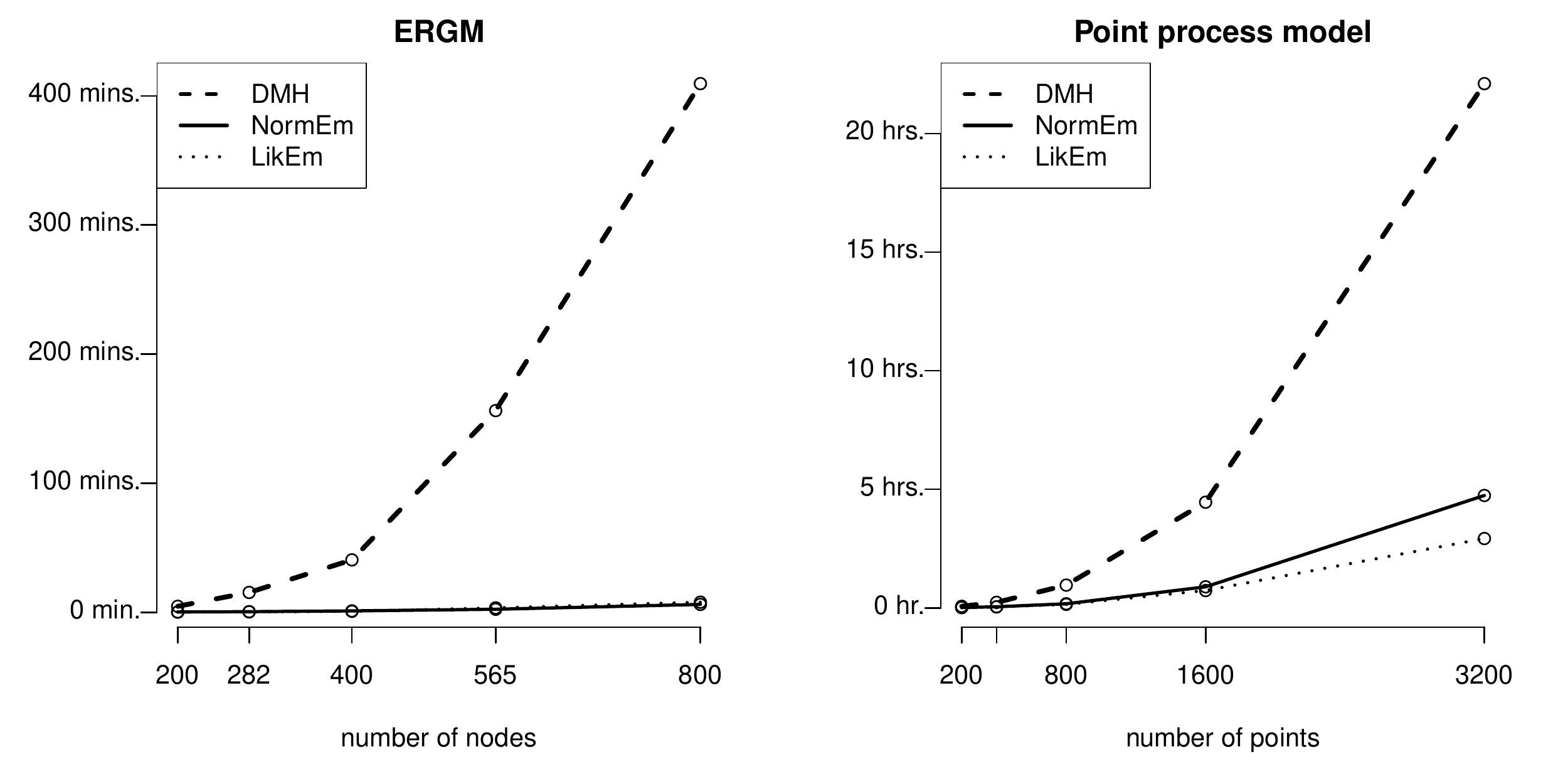}
\end{center}
\caption{Illustration of the observed computing time for algorithms. For ERGM, time is measured for simulation settings in Table~\ref{ERGMsetting} with 25,000 iterations. For point process model, time is measured for RSV-B simulation settings in \cite{goldstein2014attraction} with 40,000 iterations.}
\label{complexityfigure}
\end{figure}

Figure~\ref{complexityfigure} is the observed computing time for algorithms with different scales in both models. In the ERGM, it is observed that $\mathcal{O}(n^3)$ complexity for DMH (for $\sqrt{2}$ times larger nodes, computing time takes about $2\sqrt{2}$ times longer). Complexities of both emulation approaches are similar to each other because the unnormalized likelihood is not expensive to calculate. The results in Figure~\ref{complexityfigure} are consistent with our calculations.
 
\section{Discussion}

~~~~~In this manuscript, we have proposed fast Gaussian process-based function emulation approaches for Bayesian inference. We describe two algorithms -- one specifically targeted at doubly intractable distributions, while the other applies broadly to problems where the likelihood functions are expensive. Our study shows that our function emulation approaches provide comparable results at far lower computational cost then existing algorithms. We have also studied bounds on the total variation between a Markov chain with the exact target distribution, and the Markov chain of our approximate algorithm. 
Our study of applications to real and simulated data applications shows that our function emulation approaches provide similar results to the best current algorithms, but at a fraction of the computational cost.  

There have been a number of recent proposals for efficient precomputation approaches for intractable normalising function problems. These include the precomputation for Monte Carlo approximations in \cite{boland2017efficient}, and preprocessing for approximate Bayesian computation \citep{moores2015pre}. The precomputation step in our method is embarrassingly parallel in that the importance sampling estimates can be constructed entirely in parallel. Therefore, with relatively little effort, the computational costs can be reduced by a factor corresponding to the number of available threads. This can be helpful given the increasing availability of parallel computing resources.
 
We note that if there are irregularities, for example multimodalities, in the likelihood function, it becomes more challenging to emulate accurately. In such cases, we can extend our methods via a local Gaussian process approximation. For instance, \cite{gramacy2015local} provide a nonstationary modeling framework by constructing a Gaussian process emulator based on a local subset of the data. Their approach can provide accurate estimates in the presence of irregularities and can take advantage of parallel computation for local design. We note that it would be ideal to use a more efficient approach than importance sampling for approximating normalizing functions. However, in the interest of computational efficiency, particularly our ability to easily parallelize portions of the algorithm, as well as the limited alternative methods for approximating normalizing functions efficiently, importance sampling appears to be a good choice. Although our emulation approaches are scalable for high-dimensional data sets, we note that their performance relies on the choice of particles. Choosing particles well becomes a bigger challenge with increasing parameter dimensions. By a clever choice of particles, \cite{drovandi2018accelerating} applies a Gaussian process approximation to an 8-dimensional parameter in the pseudo-marginal context. However, for doubly-intractable distributions there are practical implementation issues for approximating $Z(\bm{\theta})$ for higher dimensional $\bm{\theta}$, as pointed out in \cite{park2018bayesian}. For example, \cite{atchade2008bayesian,liang2015adaptive} provide multiple importance sampling approaches for  robust estimates of $Z(\bm{\theta})$, but this method suffers from slow mixing of the stochastic approximation. To our knowledge  DMH is the only algorithm applicable for higher parameter dimension doubly-intractable problems. However DMH does not provide likelihood estimates as a by-product, because the auxiliary variables introduced by the algorithm result in $Z(\bm{\theta})$ being canceled out in the Metropolis-Hastings acceptance ratio (see Step 3 in the Algorithm 3). Furthermore the DMH algorithm requires simulation of a high-dimensional auxiliary variable with each iteration, which is computationally demanding for the problems considered here. The methods we describe in this paper are effective for low-dimensional parameter spaces, involving high-dimensional data sets. Inference for doubly-intractable distributions arising from high-dimensional parameter space models with large data sets (e.g. thousands of nodes with 10 dimensional parameter space) remains an open challenge. Hence, our methods are ideally suited to parameter dimensions similar to those we considered in our examples, that is, between 1 and 4. Given our current computing resources we find that our methods are well suited to problems where simulating auxiliary variables, that is, producing a single draw from the probability model $h(\mathbf{x}|\bm{\theta})$, takes under 20 seconds. As our examples in Section 4 illustrate, this allows for our method to be applicable to several classes of practical models for which current methods are infeasible. However, an open question is how to extend the algorithms to work beyond these parameter dimensions. Our methods can be extended to moderate dimensional parameter space models, but the number of particles would then increase exponentially with dimension, which slows computing. Therefore in this manuscript we choose particles carefully by using ABC or a short run of DMH \cite[see also][]{atchade2008bayesian,liang2015adaptive}. There are function interpolation approaches that add design points sequentially to improve the accuracy of approximation \citep[cf.][]{joseph2012bayesian,joseph2015sequential,conrad2016accelerating, wang2017adaptive}. However, direct application of these methods is challenging because they require sequential optimization, which is computationally expensive  as there are no robust estimates for high-dimensional $Z(\bm{\theta})$ in the doubly-intractable distribution context. Developing extensions of our approach to high-dimensional parameter models may provide an interesting avenue for future research.

\section*{Acknowledgement}
MH and JP were partially supported by the National Science Foundation through NSF-DMS-1418090. The authors are grateful to Galin Jones, Omiros Papaspiliopoulos and Alexander Mitrophanov for helpful discussions. 

\section*{Supplementary Material}
Supplementary material available online contains proofs, complexity calculations, and details for the model for infectious disease dynamics briefly described in Section 4.3 . It also provides tables and figures not included in the manuscript.

\clearpage

\appendix
\begin{center}
\title{\LARGE\bf Appendix}\\~\\
\end{center}

\section{Proof of Theorem~1}\label{GPproofapp}
Consider a target distribution $\pi(\bm{\theta}|\mathbf{x})$ whose Markov chain transition kernel is $\mathbf{P}$. The acceptance probability of which is 
\begin{equation}
\alpha(\bm{\theta},\bm{\theta}') = \frac{q(\bm{\theta}|\bm{\theta}')p(\bm{\theta}')h(\mathbf{x}|\bm{\theta}')Z(\bm{\theta})}{q(\bm{\theta}'|\bm{\theta})p(\bm{\theta})h(\mathbf{x}|\bm{\theta})Z(\bm{\theta}')}.
\end{equation}

$\widehat{Z}_{IS}(\bm{\theta})$ and $\widehat{Z}_{GP}(\bm{\theta})$ denote an importance sampling estimate and a Gaussian process arppoximation as in \eqref{IS} and \eqref{BLUP1} respectively. By plugging in the $\widehat{Z}_{IS}(\bm{\theta})$ into the acceptance probability, first-stage approximated transition kernel $\mathbf{\widehat{P}}_{IS}$ can be constructed; the acceptance probability of which is $\widehat{\alpha}_{IS}(\bm{\theta},\bm{\theta}')$. Second-stage approximated kernel $\mathbf{\widehat{P}}_{GP}$ is constructed by replacing $\widehat{Z}_{IS}(\bm{\theta})$ with  $\widehat{Z}_{GP}(\bm{\theta})$ and $\widehat{\alpha}_{GP}(\bm{\theta},\bm{\theta}')$ is the corresponding acceptance probability.

\subsection{Approximation Error of Importance Sampling Estimates}

Bound of difference between the acceptance probabilities of $\mathbf{P}$ and $\mathbf{\widehat{P}}_{IS}$ can be derived as follows. 
\begin{equation}
\begin{split}
|\widehat{\alpha}_{IS}(\bm{\theta},\bm{\theta}')-\alpha(\bm{\theta},\bm{\theta}')| & =\frac{q(\bm{\theta}|\bm{\theta}')p(\bm{\theta}')h(\mathbf{x}|\bm{\theta}')}{q(\bm{\theta}'|\bm{\theta})p(\bm{\theta})h(\mathbf{x}|\bm{\theta})}\left|\frac{\frac{1}{N}\sum_{l}\frac{h(\mathbf{x}_{l}|\bm{\theta})}{h(\mathbf{x}_{l}|\bm{\widetilde{\theta}})}}{\frac{1}{N}\sum_{l}\frac{h(\mathbf{x}_{l}|\bm{\theta}')}{h(\mathbf{x}_{l}|\bm{\widetilde{\theta}})}}-\frac{Z(\bm{\theta})}{Z(\bm{\theta}')}\right|\\
& \leq \epsilon(N)\frac{q(\bm{\theta}|\bm{\theta}')p(\bm{\theta}')h(\mathbf{x}|\bm{\theta}')}{q(\bm{\theta}'|\bm{\theta})p(\bm{\theta})h(\mathbf{x}|\bm{\theta})} \leq \epsilon(N)c_{p}^2c_{q}^2\frac{K}{k}~~~\text{a.s.}\\
\end{split}
\end{equation}

The first inequality is from the ergodic theorem and continuous mapping theorem. With increasing $N$, importance sampling estimates converge to $Z(\bm{\theta})$ almost surely. The second inequality is from Assumption~1-3 in the Theorem~1.

We now show that $\mathbf{P}$ is uniformly ergodic for measurable subset $B$ of $\bm{\Theta}$. From Assumption~1-3 in the Theorem~1, Markov transition kernel may be bounded as follows,  
\begin{equation}
\begin{split}
\mathbf{P}(\bm{\theta},B) & = 
\int_{B}\delta_{\bm{\theta}}(d\bm{\theta}')[1-\int d\mathbf{t}q(\mathbf{t}|\bm{\theta})\min\{1,\alpha(\bm{\theta},\mathbf{t})\}] + \int_{B}d\bm{\theta}'q(\bm{\theta}'|\bm{\theta})\min\{1,\alpha(\bm{\theta},\bm{\theta}')\}\\
& \geq \int_{B}d\bm{\theta}'q(\bm{\theta}'|\bm{\theta})\min\{1,\alpha(\bm{\theta},\bm{\theta}')\}\\
& \geq \frac{k^2}{c_{p}^2c_{q}^2K^2}\int_{B}d\bm{\theta}'q(\bm{\theta}'|\bm{\theta})\\
& \geq \frac{k^2}{c_{p}^2c_{q}^3K^2}\int_{B}d\bm{\theta}'.
\end{split}
\end{equation}

According to Theorem 16.0.2 and Theorem 16.2.4 in \cite{meyn1993markov}, this proves that $\sup_{\bm{\theta}}\|\delta_{\bm{\theta}}\mathbf{P}-\pi(.)\| \leq C\rho^{n}$, where $C=2$, $\rho=1-k^2/(c_{p}^3c_{q}^3K^2)$ and $0<\rho<1$. Hence, the following conditions hold: (1) the difference between acceptance probabilities is bounded, and (2) the Markov chain with transition kernel $\mathbf{P}$ is uniformly ergodic. The assumptions of Corollary 2.3 in \cite{alquier2014noisy} are therefore satisfied which implies that the approximation error of importance sampling estimates is 
\begin{equation}
\label{part1}
\|\delta_{\bm{\theta}_{0}}\mathbf{P}^{n}-\delta_{\bm{\theta}_{0}}\mathbf{\widehat{P}}_{IS}^{n} \| \leq \epsilon(N)c_{p}^2c_{q}^2\frac{K}{k}(\lambda+\frac{C \rho^{\lambda}}{1-\rho})~~~\text{a.s.},~~~\lambda=\ceil*{\frac{\log(1/C)}{\log(\rho)}}.
\end{equation}

\subsection{Approximation Error of Gaussian Process Emulation}

Now we consider the second-stage approximation of the Metropolis-Hastings acceptance probability. Similar to the previous section, we can derive bound of difference between acceptance probabilities of $\mathbf{\widehat{P}}_{GP}$ and $\mathbf{\widehat{P}}_{IS}$. The difference between acceptance probabilities is 
\begin{equation}
|\widehat{\alpha}_{GP}(\bm{\theta},\bm{\theta}')-\widehat{\alpha}_{IS}(\bm{\theta},\bm{\theta}')|=\frac{q(\bm{\theta}|\bm{\theta}')p(\bm{\theta}')h(\mathbf{x}|\bm{\theta}')}{q(\bm{\theta}'|\bm{\theta})p(\bm{\theta})h(\mathbf{x}|\bm{\theta})}\left|
\frac{\widehat{Z}_{GP}(\bm{\theta})}{\widehat{Z}_{GP}(\bm{\theta}')}-\frac{\frac{1}{N}\sum_{l}\frac{h(\mathbf{x}_{l}|\bm{\theta})}{h(\mathbf{x}_{l}|\bm{\widetilde{\theta}})}}{\frac{1}{N}\sum_{l}\frac{h(\mathbf{x}_{l}|\bm{\theta}')}{h(\mathbf{x}_{l}|\bm{\widetilde{\theta}}}}\right|.
\end{equation}

Since the parameter space $\bm{\Theta}$ is assumed to be compact, there exists a finite $d$-number of open balls with radius $r(d)$ that can cover $\bm{\Theta}$. Let $( \bm{\theta}^{(1)},...\bm{\theta}^{(d)} )$ be center of the $d$-balls respectively. Here, $\widehat{Z}(\bm{\theta})_{GP}$ is a continuous function with respect to $\bm{\theta}$. This is because $\widehat{Z}_{GP}(\bm{\theta})$ is a linear function of $\widehat{Z}_{IS}(\bm{\theta})$, which is a continuous function of $\bm{\theta}$. This satisfies 
\begin{equation}
\forall r(d)>0~~\exists ~ \epsilon(r(d))>0~~\text{s.t.}~~\left|\widehat{Z}_{GP}(\bm{\theta})-\frac{1}{N}\sum_{l}\frac{h(\mathbf{x}_{l}|\bm{\theta})}{h(\mathbf{x}_{l}|\bm{\widetilde{\theta}})}\right|< \epsilon(r(d)),
\end{equation}
for every $\bm{\theta} \in \bm{\Theta}$ (i.e. uniformly convergent). $\widehat{Z}_{GP}(\bm{\theta})/\widehat{Z}_{IS}(\bm{\theta})$ is also continuous and has value $1$, when $\bm{\theta}=\bm{\theta}^{(j)}$ for $j=1,...,d$. Therefore with continuous mapping theorem and Assumption~1-3 in the Theorem~1, the difference in acceptance probability approximations may be bounded as
\begin{equation}
|\widehat{\alpha}_{GP}(\bm{\theta},\bm{\theta}')-\widehat{\alpha}_{IS}(\bm{\theta},\bm{\theta}')| \leq \epsilon(d)\frac{q(\bm{\theta}|\bm{\theta}')p(\bm{\theta}')h(\mathbf{x}|\bm{\theta}')}{q(\bm{\theta}'|\bm{\theta})p(\bm{\theta})h(\mathbf{x}|\bm{\theta})} \leq \epsilon(d)c_{p}^2c_{q}^2\frac{K}{k}.
\end{equation}

We now show that $\mathbf{\widehat{P}}_{IS}$ is uniformly ergodic for measurable subset $B$ of $\bm{\Theta}$. From Assumption~1-3 in Theorem~1, we obtain, 
\begin{equation}
\widehat{\alpha}_{IS}(\bm{\theta},\bm{\theta}')=\frac{q(\bm{\theta}|\bm{\theta}')p(\bm{\theta}')h(\mathbf{x}|\bm{\theta}')\frac{1}{N}\sum_{l}\frac{h(\mathbf{x}_{l}|\bm{\theta})}{h(\mathbf{x}_{l}|\bm{\widetilde{\theta)}}}}
{q(\bm{\theta}'|\bm{\theta})p(\bm{\theta})h(\mathbf{x}|\bm{\theta})\frac{1}{N}\sum_{l}\frac{h(\mathbf{x}_{l}|\bm{\theta}')}{h(\mathbf{x}_{l}|\bm{\widetilde{\theta)}}}}\geq \frac{k^3}{c_{p}^2c_{q}^2K^{3}}.
\label{ISacc}
\end{equation}

Therefore, the importance sampling approximation of the original Markov transition kernel is bounded as follows,
\begin{equation}
\begin{split}
\mathbf{\widehat{P}}_{IS}(\bm{\theta},B) & = 
\int_{B}\delta_{\bm{\theta}}(d\bm{\theta}')[1-\int d\mathbf{t}q(\mathbf{t}|\bm{\theta})\min\{1,\widehat{\alpha}_{IS}(\bm{\theta},\mathbf{t})\}] + \int_{B}d\bm{\theta}'q(\bm{\theta}'|\bm{\theta})\min\{1,\widehat{\alpha}_{IS}(\bm{\theta},\bm{\theta}')\}\\
& \geq \int_{B}d\bm{\theta}'q(\bm{\theta}'|\bm{\theta})\min\{1,\widehat{\alpha}_{IS}(\bm{\theta},\bm{\theta}')\}\\
& \geq \frac{k^3}{c_{p}^2c_{q}^2K^3}\int_{B}d\bm{\theta}'q(\bm{\theta}'|\bm{\theta})\\
& \geq \frac{k^3}{c_{p}^2c_{q}^3K^3}\int_{B}d\bm{\theta}',
\end{split}
\end{equation}

where the second inequality follows from \eqref{ISacc}. According to the Theorem 16.0.2 and Theorem 16.2.4 in \cite{meyn1993markov}, this proves that $\sup_{\bm{\theta}}\|\delta_{\bm{\theta}}\mathbf{\widehat{P}}_{IS}-\widehat{\pi}_{IS}(.)\| \leq C\rho^{n}$, where $C=2$, $\rho=1-k^3/(c_{p}^3c_{q}^3K^3)$ and $0<\rho<1$. In a similar fashion to the previous section, the assumptions of Corollary 2.3 in \cite{alquier2014noisy} are therefore satisfied which implies that the approximation error of Gaussian process approximations is

\begin{equation}
\label{part2}
\|\delta_{\bm{\theta}_{0}}\mathbf{\widehat{P}}_{IS}^{n}-\delta_{\bm{\theta}_{0}}\mathbf{\widehat{P}}_{GP}^{n} \| \leq \epsilon(d)c_{p}^2c_{q}^2\frac{K}{k}(\lambda+\frac{C \rho^{\lambda}}{1-\rho}),~~~\lambda=\ceil*{\frac{\log(1/C)}{\log(\rho)}}.
\end{equation}

\subsection{Approximation Error of the normalising Function Emulation Approach}
From \eqref{part1} and \eqref{part2}, and ignoring the constants $c_{p}^2c_{q}^2\frac{K}{k}(\lambda+\frac{C \rho^{\lambda}}{1-\rho})$, the approximation error for the normalising function emulation approach is
\begin{equation}
\begin{split}
\|\pi(.)-\delta_{\bm{\theta}_{0}}\mathbf{\widehat{P}}_{GP}^{n}\| & \leq 
\|\pi(.)-\delta_{\bm{\theta_{0}}}\mathbf{P}^{n}\| + \|\delta_{\bm{\theta_{0}}}\mathbf{P}^{n}-\delta_{\bm{\theta}_{0}}\mathbf{\widehat{P}}_{IS}^{n} \| + \|\delta_{\bm{\theta_{0}}}\mathbf{\widehat{P}}_{IS}^{n}-\delta_{\bm{\theta}_{0}}\mathbf{\widehat{P}}_{GP}^{n}\|\\
& \leq C\rho^{n} + \epsilon(N) + \epsilon(d)~~~\text{a.s.}
\end{split}
\end{equation}

\section{Proof of Corollary~1}\label{GPproofapp2}

By plugging in the $\widehat{L}_{IS}(\bm{\theta}|\mathbf{x})$ into the acceptance probability, first-stage approximated transition kernel $\mathbf{\widehat{P}}_{IS}$ can be constructed; the acceptance probability of which is $\widehat{\alpha}_{IS}(\bm{\theta},\bm{\theta}')$. Second-stage approximated kernel $\mathbf{\widehat{P}}_{GP}$ is constructed by replacing $\widehat{L}_{IS}(\bm{\theta}|\mathbf{x})$ with  $\widehat{L}_{GP}(\bm{\theta}|\mathbf{x})$ and $\widehat{\alpha}_{GP}(\bm{\theta},\bm{\theta}')$ is the corresponding acceptance probability. 

\subsection{Approximation Error of Importance Sampling Estimates}

This part is identical to the proof in a normalising function emulation. The total variation norm distance between the true kernel and the first-stage approximated transition kernel is derived as \eqref{part1}.

\subsection{Approximation Error of Gaussian Process Emulation}

The difference between acceptance probabilities is 
\begin{equation}
|\widehat{\alpha}_{GP}(\bm{\theta},\bm{\theta}')-\widehat{\alpha}_{IS}(\bm{\theta},\bm{\theta}')|=\frac{q(\bm{\theta}|\bm{\theta}')p(\bm{\theta}')}{q(\bm{\theta}'|\bm{\theta})p(\bm{\theta})}\left|
\frac{\widehat{L}_{GP}(\bm{\theta}'|\mathbf{x})}{\widehat{L}_{GP}(\bm{\theta}|\mathbf{x})}-\frac{\widehat{L}_{IS}(\bm{\theta}'|\mathbf{x})}{\widehat{L}_{IS}(\bm{\theta}|\mathbf{x})}\right|.
\end{equation}

In a similar fashion to the previous section, we obtain, 
\begin{equation}
\forall r(d)>0~~\exists ~ \epsilon(r(d))>0~~\text{s.t.}~~\left|\widehat{L}_{GP}(\bm{\theta}|\mathbf{x})- \widehat{L}_{IS}(\bm{\theta}|\mathbf{x})  \right|< \epsilon(r(d)),
\end{equation}
for every $\bm{\theta} \in \bm{\Theta}$ (i.e. uniformly convergent) by using a compact parameter space assumption. Therefore with continuous mapping theorem and Assumption~1-3 in the Corollary~1, bound of difference between acceptance probabilities of $\mathbf{\widehat{P}}_{GP}$ and $\mathbf{\widehat{P}}_{IS}$ can be derived as 
\begin{equation}
|\widehat{\alpha}_{GP}(\bm{\theta},\bm{\theta}')-\widehat{\alpha}_{IS}(\bm{\theta},\bm{\theta}')| \leq \epsilon(d)\frac{q(\bm{\theta}|\bm{\theta}')p(\bm{\theta}')}{q(\bm{\theta}'|\bm{\theta})p(\bm{\theta})} \leq \epsilon(d)c_{p}^2c_{q}^2.
\end{equation}

Moreover, $\mathbf{\widehat{P}}_{IS}$ is uniformly ergodic for measurable subset $B$ of $\bm{\Theta}$ as we show in Section~A.2. The assumptions of Corollary 2.3 in \cite{alquier2014noisy} are therefore satisfied which implies that the approximation error of Gaussian process approximations is 
\begin{equation}
\label{part22}
\|\delta_{\bm{\theta}_{0}}\mathbf{\widehat{P}}_{IS}^{n}-\delta_{\bm{\theta}_{0}}\mathbf{\widehat{P}}_{GP}^{n} \| \leq \epsilon(d)c_{p}^2c_{q}^2(\lambda+\frac{C \rho^{\lambda}}{1-\rho}),~~~\lambda=\ceil*{\frac{\log(1/C)}{\log(\rho)}},
\end{equation}
where $C=2$, $\rho=1-k^3/(c_{p}^3c_{q}^3K^3)$ and $0<\rho<1$
\subsection{Approximation Error of the Likelihood Function Emulation Approach}
From \eqref{part1} and \eqref{part22}, and ignoring the constants $c_{p}^2c_{q}^2\frac{K}{k}(\lambda+\frac{C \rho^{\lambda}}{1-\rho})$, the approximation error for the likelihood function emulation approach is
\begin{equation}
\begin{split}
\|\pi(.)-\delta_{\bm{\theta}_{0}}\mathbf{\widehat{P}}_{GP}^{n}\| & \leq 
\|\pi(.)-\delta_{\bm{\theta_{0}}}\mathbf{P}^{n}\| + \|\delta_{\bm{\theta_{0}}}\mathbf{P}^{n}-\delta_{\bm{\theta}_{0}}\mathbf{\widehat{P}}_{IS}^{n} \| + \|\delta_{\bm{\theta_{0}}}\mathbf{\widehat{P}}_{IS}^{n}-\delta_{\bm{\theta}_{0}}\mathbf{\widehat{P}}_{GP}^{n}\|\\
& \leq C\rho^{n} + \epsilon(N) + \epsilon(d)~~~\text{a.s.}
\end{split}
\end{equation}

\bibliography{Reference}
\end{document}